\documentclass{ws-procs9x6-cpt22}
\begin{document}

\title{Calculations in Lorentz-breaking scalar QED}

\author{Jean C. C. Felipe,$^1$ A. Yu Petrov,$^{2}$, A. P. B. Scarpelli $^3$ and L. C. T. Brito $^4$}

\address{$^1$Instituto de Engenharia, Ci\^{e}ncia e Tecnologia, Universidade Federal dos Vales do Jequitinhonha e Mucuri, Avenida Um, 4050 - 39447-790 - Cidade Universit\'{a}ria - Jana\'{u}ba - MG- Brazil}

\address{$^2$Departamento de F\'{i}sica, Universidade Federal da Paraíba, Caixa Postal 5008,
	58051-970 Jo\~{a}o Pessoa, Para\'{i}ba, Brazil}

\address{$^3$Centro Federal de Educa\c{c}\~ao Tecnol\' ogica - MG - MG Avenida Amazonas, 7675 - 30510-000 - Nova Gameleira - Belo Horizonte -MG - Brazil}

\address{$^4$Departamento de F\'{i}sica, Universidade Federal de Lavras, Caixa Postal 3037,
	37200-000, Lavras, MG, Brasil}


\begin{abstract}
We purpose a study of CPT-even Lorentz-breaking extension of the scalar QED. We calculate the one-loop contributions in the Lorentz-violating parameters to the two-point functions of
scalar and gauge fields. We found that the two background tensors, coming from the two sectors (scalar and gauge) are mixed in the one-loop corrections. This shows that these two Lorentz-breaking terms cannot be studied in an isolated form. Besides, the results in the gauge sector are confirmed to be transversal.
\end{abstract}

\bodymatter

\section{Introduction}

The formulation of the Lorentz-violating (LV) extension of Standard Model (SME) brought interest to perturbative calculations in several Lorentz symmetry breaking models \cite{Colladay1997}. So, one of the motivations for such calculations consists in the development of a scheme from which the desired LV terms arise as quantum corrections to some fundamental theory where vector (gauge) and scalar fields are coupled to some spinor field \cite{Jackiw1999}. On the experimental side, the search for possible effects of Lorentz violation may present imprints of quantum gravity at low energy \cite{Kost1999}. In particular, from the perspective of the Standard Model Extension (SME) program, proposed tests involve, for example, gravitational waves \cite{Sch2017}. The coefficients in the SME are very small, they cannot be disregarded. In particular, for the dimensionless coefficients in which we are interested in this work, we have that $c^{\mu\nu}$ (scalar sector) is about $10^{-16}$ and $\kappa^{\mu\nu\lambda\rho}$ (photon sector) is about $10^{-17}$ \cite{arxiv}.

\section{The model}
\begin{figure}[!h]
\begin{minipage}[!h]{0.45\linewidth}
\includegraphics[width=\linewidth]{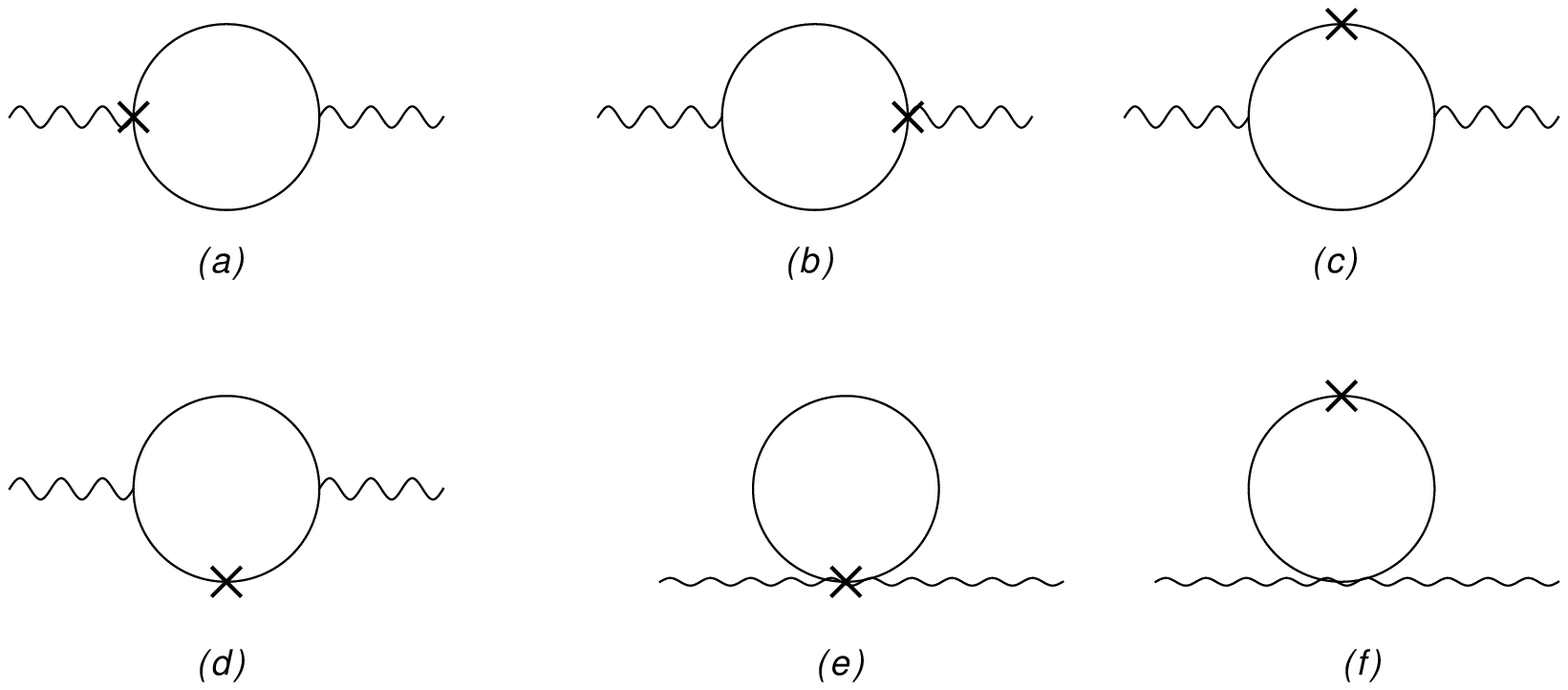}
\caption{Diagrammatic representation of the one-loop two-point function of the gauge field at first-order.}
\label{Fig1}
\end{minipage} \hfill
\begin{minipage}[!h]{0.45\linewidth}
\includegraphics[width=\linewidth]{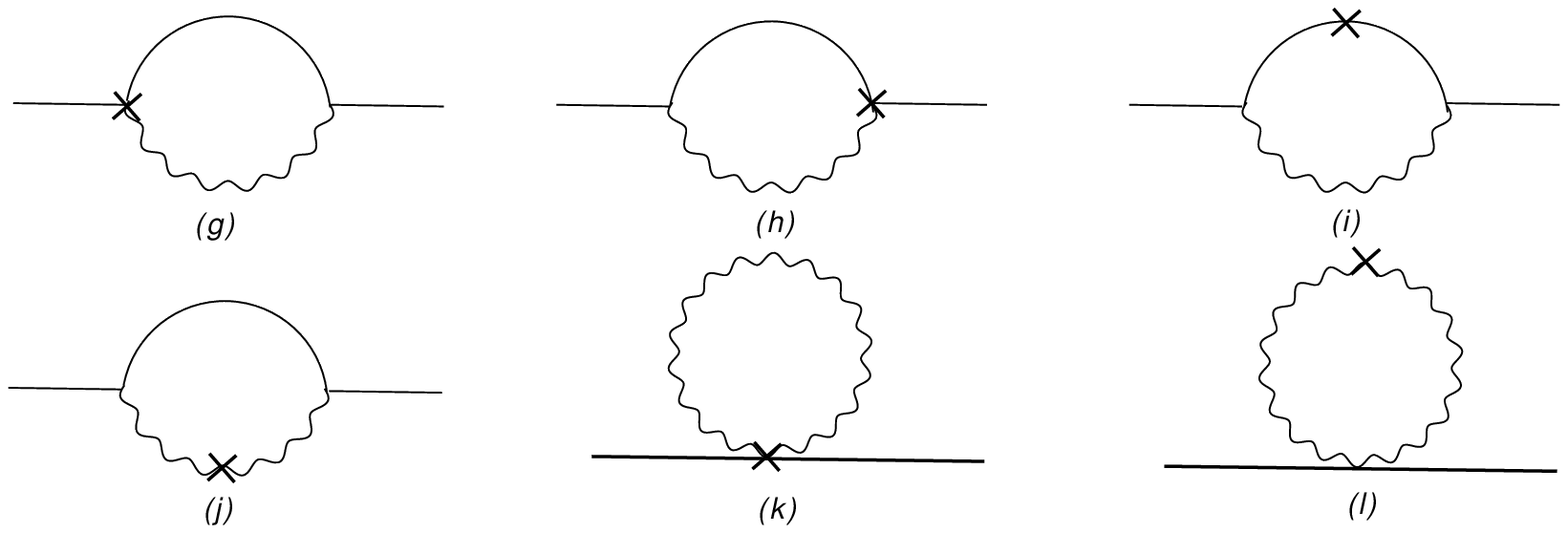}
\caption{Diagrammatic representation of the scalar field two-point function at first-order.}
\label{Fig2}
\end{minipage}
\end{figure}

We start with the CPT-even QED with the Lagrangian
\begin{eqnarray}
{\cal L}=(D_{\mu}\phi)^*(\eta^{\mu\nu}+c^{\mu\nu})(D_{\nu}\phi)-m^2\phi^*\phi-\frac{1}{4}F_{\mu\nu}F^{\mu\nu}+\frac{1}{4}\kappa_{\mu\nu\lambda\rho}F^{\mu\nu}F^{\lambda\rho}
\end{eqnarray}
Here, the Lorentz symmetry breaking is introduced both in scalar and gauge sectors through additive terms proportional to the background tensors $c^{\mu\nu}$ (symmetric and traceless) and $\kappa^{\mu\nu\lambda\rho}$ (has the same symmetry as the Riemann curvature tensor). We would like to study the one-loop corrections to the two point functions both of the scalar and gauge sectors. We need to calculate the one loop contributions depicted by diagrmas on Figures \ref{Fig1} and \ref{Fig2}, respectively (the wavy and solid lines represent the photon and scalar propagators, respectively, and the crosses indicate insertions of LV parameters).

\section{Results and Discussion}

We calculated explicitly the contributions from all diagrms depcted. In order to do this, we used the Implicit Regularization framework \cite{IReg2022,Jean2022}. We found the following result for the vacuum polarization tensor for the scalar sector

\begin{eqnarray}
	\Pi^{\mu\nu}(p^2,m^2)&=&\frac{1}{3}\left\{p^2c^{\mu\nu}-p_{\alpha}(c^{\alpha\mu}p^{\nu}+c^{\alpha\nu}p^{\mu})+\eta^{\mu\nu}c^{\alpha\beta}p_{\alpha}p_{\beta}
	\right\}\times\nonumber\\ &\times&
	\left\{I_{log}(m^2) - \frac{b}{p^2}\left[(p^2-4 m^2)Z_0(p^2,m^2)-\frac{2}{3}p^2\right]\right\} + \nonumber\\
	&+&\frac{b}{3} c^{\alpha\beta}\frac{p_{\alpha}p_{\beta}}{p^4} (p^{\mu}p^{\nu} - p^2\eta^{\mu\nu})[6m^2 Z_0(p^2,m^2) + p^2],
\end{eqnarray}
in which $b=\frac{i}{(4\pi)^2}$
and
\begin{equation}
	\Pi(p^2,m^2)=2p_{\alpha}p_{\beta}c^{\alpha\beta}\Big\{-2I_{log}(m^2)+b\Big[\tilde{Z_{0}}+3\tilde{Z_{2}}+2p^{2}(\tilde{Y_{3}}-\tilde{Y_{4}})\Big]\Big\}
\end{equation}
for the matter sector (for more details about the functions $Z_{k}(p^{2},m^{2}$ and $\tilde{Y}_{k}(p^{2},m^{2})$, see \cite{IReg2022}).

The complete result for the scalar field self-energy, given by the sum of the contributions on $c^{\mu\nu}$ and $\kappa^{\mu\nu\alpha\beta}$, looks like
\begin{eqnarray}
	\label{complete}
	\Pi(p^2,m^2)&=& 2p_{\alpha}p_{\beta}c^{\alpha\beta}\Big\{-2I_{log}(m^2)+b\Big[\tilde{Z_{0}}+3\tilde{Z_{2}}+2p^{2}(\tilde{Y_{3}}-\tilde{Y_{4}})\Big]\Big\} +
	\nonumber\\ &-&
	\frac{1}{4}p_{\alpha}p_{\beta}\kappa^{\alpha\,\,\,\gamma\beta}_{\,\,\,\gamma}\Big\{I_{log}(m^{2})-2b\tilde{Z_{1}}\Big\}.
\end{eqnarray}

\section{Concluding Remarks}

We started with the simplest CPT-even Lorentz-violating extension of scalar QED, where LV parameters are introduced both in scalar and gauge sectors. We found that aether-like structures arise in both cases. While our results are divergent, their renormalization can be performed through simple wave function renormalizations for aether terms in both sectors. The mainly results of this work have been published in \cite{Jean2022}.

\section{Acknowledgments}

A. P. B. Scarpelli and A. Yu Petrov acknowledges CNPq by financial support.. Jean C. C. Felipe acknowledges the UFVJM by support.

\end{document}